# Electrons and holes in Si quantum well: a room-temperature transport and drag resistance study

M. Prunnila*, S. J. Laakso, J. M. Kivioja, and J. Ahopelto

VTT Micro and Nanoelectronics, P.O.Box 1000, FI-02044 VTT, Espoo, FINLAND

We investigate carrier transport in a single 22 nm-thick double-gated Si quantum well device, which has independent contacts to electrons and holes. Conductance, Hall density and Hall mobility are mapped in a broad double-gate voltage window. When the gate voltage asymmetry is not too large only either electrons or holes occupy the Si well and the Hall mobility shows the fingerprints of volume inversion/accumulation. At strongly asymmetric double-gate voltage an electric field induced electron-hole (EH) bi-layer is formed inside the well. The EH drag resistance $R_{he}$ is explored at balanced carrier densities: $R_{he}$ decreases monotonically from 860 Ω to 37 Ω when the electron and hole density is varied between ~0.4−1.7x$10^{16}$ m$^{-2}$.

*mika.prunnila@vtt.fi

Silicon-on-insulator (SOI) based double-gate $SiO_2$-Si-$SiO_2$ quantum well field-effect-transistors (FET) [see Fig. 1 (a)] are being actively investigated.[1,2] The double-gate geometry provides means to effectively adjust the total carrier density inside the FET channel, which leads, e.g., to suppression of the fatal short channel effects and enhancement of the sub-threshold characteristics. The double-gate structure also enables adjustment of the width and position of the carrier distribution (or wave functions) inside the Si well. Carrier gas can be tuned to the middle of the well at symmetric double-gate voltage [see Fig. 1 (b)] or pushed against either of the Si-$SiO_2$ interfaces by applying an asymmetric gate voltage, which has a relatively large effect on the carrier mobility.[3,4] The high dielectric strength of the $SiO_2$ also enables strongly asymmetric gate voltage that can overcome the energy gap of Si even in relatively thin Si wells. This leads to formation of a field induced electron-hole (EH) bi-layer inside the device [see Fig. 1. (c)] analogously to the III-V system of Ref. [5], but without any heterostructure based energy barrier between the different layers.

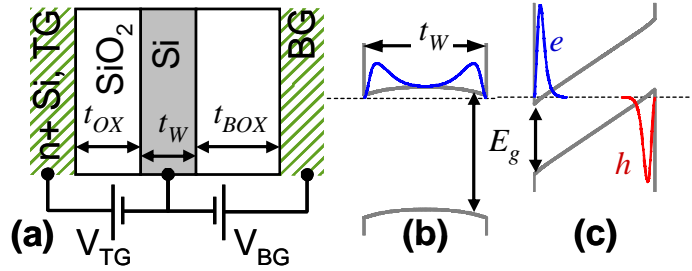

**Fig. 1:** (Color online) (a) Schematic cross-section of a double gated Si quantum well together with external top gate (TG) and back gate (BG) voltage sources. (b)&(c) Self-consistently calculated quantum mechanical electron (blue) and hole (red) distributions at different gate voltages within the Hartree approximation. The energy gap $E_g$ = 1.12 eV between the conduction and valence bands defines the energy scales. 1 eV corresponds to carrier density of $5 \times 10^{24}$ $m^{-3}$. The dashed line is the chemical potential and well thickness is $t_W$ = 22 nm. In (b) the gate voltage is positive and symmetric ($V_{BG} = V_{TG}t_{BOX}/t_{OX}$) with total 2D electron density of $2 \times 10^{16}$ $m^{-2}$. In (c) the gate voltage is asymmetric with $V_{BG} \sim -V_{TG}t_{BOX}/t_{OX}$ and the 2D electron and hole densities are $1.214 \times 10^{16}$ $m^{-2}$.

In this Letter, we investigate room temperature transport properties of a 22 nm-thick double-gate Si well. The device has n-type and p-type contact regions enabling simultaneous measurements on electrons and holes. Conductance, Hall density and Hall mobility are mapped in a broad double-gate voltage window. When the gate voltage asymmetry is not too large only either electrons or holes occupy the Si well. In the case of electrons we reproduce the results of Ref. [4] and for the holes we find qualitatively similar gate voltage dependency of mobility, but with significantly lower magnitude consistent with Ref. [6]. At strongly asymmetric double-gate voltage we demonstrate an EH bi-layer: conductance, density and mobility are simultaneously non-zero for both carrier types. This EH system enables investigation of inter-layer friction, which has been a topical subject since the early drag experiments on electron-electron [7] and EH [8] bi-layers (see Ref. [9] for a review). Here we demonstrate an EH drag resistance measurement at equal electron and hole densities.

The Si quantum well devices were fabricated on commercially available bonded SOI wafers. The fabrication process follows closely the procedures described in Ref. [10]. The major difference to standard SOI FET is that instead of having only n+ contacts the present devices also have p+ contact regions, which provide galvanic contact to the hole gas inside the 22 nm-thick Si well. The top (back) gate oxide thickness is $t_{OX}$ = 50 nm ($t_{BOX}$ = 83 nm). Figure 2 shows an optical micrograph of the Hall bar device explored in this work. Biasing current is applied independently into the each carrier system from the ends of the bar with a p+/n+-finger structure that is illustrated in Fig. 2(b). As in standard Si FET fabrication the p+ and n+ regions are self-aligned with respect to the top gate. The contact between electron (hole) gas inside the Si well and p+ (n+) regions is a pn-junction with high isolation: the small bias resistance of these junctions was found to be above 10 GΩ.

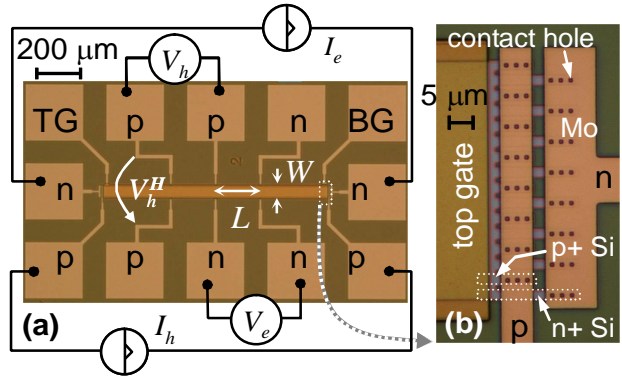

**Fig. 2:** (Color online) (a) Optical microscope image of the lateral device geometry utilized in the experiments and illustration of the EH biasing scheme. Hall voltage of electrons is omitted for the sake of clarity. Letters p and n on the Mo pads refer to the type of Si to which the particular pad is connected. The width of the channel W = 50 μm and the probe distance L = 200 μm. (b) Magnification of the current contacts illustrating the p+/n+-finger structure. Dimensions of one n-type and one p-type finger is illustrated: the dotted rectangles depict holes in the implantation masks. The undoped Si well resides below the top gate. The areas outside the active Si region are $SiO_2$.

The electronic properties were investigated at room temperature in the linear response regime, when the EH system is described by a set of linear equations $V_i = (W/L)G_{ij}I_j$, where $I_i$ and $V_i$ are the current and the voltage of the different type of carriers as depicted in Fig. 2 (a) (*i,j* = *e* for electrons and *i,j* = *h* for holes). The elements $G_{ij}$ define the conductance matrix and its inversion gives the resistance matrix with elements $R_{ij}$. The intra-layer transport experiments were performed by utilizing standard lock-in techniques. The Hall voltage $V_i^H$ was measured at 2-4 magnetic flux density values below |*B*| = 3 T and the Hall carrier density $n_i^H$ was given by the slope $(I_i/e)\Delta B/\Delta V_i^H$. Hall mobility $\mu_i^H$ was calculated from relation $G_{ii} = en_i^H \mu_i^H$. Due to high resistance of the hole system and the strong capacitive coupling between electrons and holes we adopted a DC technique in the extraction of the drag resistance $R_{he}$: 21 electron layer current $I_e$ values were applied and resulting hole layer voltage $V_e$ was measure (the voltage over the whole electron layer was always kept below 10 mV). Linear fit to the obtained data set gave $R_{he}$.

The diagonal elements $G_{ee}$ and $G_{hh}$ as a function of the top ($V_{TG}$) and back gate ($V_{BG}$) voltages are shown in Fig. 3(a). Figures 3(b) and 3(c) show Hall carrier densities and mobilities, respectively. Conductance, Hall density and mobility of electron and hole systems behave similarly (provided that one inverts $V_{TG}$ and $V_{BG}$). The major discrepancy arises from the typical low mobility of holes consistent with Ref. [6]. Note that at high hole densities and close to $V_{TG(BG)} \sim 0$ there is a gate voltage region where the slope $\partial n_i^H / \partial V_{TG(BG)} \sim 0$. The feature can be observed also in $n_e^H$ as a small change in the slope. Comparison with the total carrier density $n_i$ contours of Ref. [4] suggests that $n_i^H$ contours are distorted from $n_i$ when the density is high and $V_{TG,BG} < 0$ or $V_{TG,BG} > 0$. A similar effect in low temperature Hall density of a 16.5 nm-thick Si well has been addressed to appearance of carrier species with different mobilities.[11] This interpretation most likely fits to the present data also: at high electron or hole density the effective energy barrier in the middle of the well [see Fig 1(b) for electrons] increases and two separate 2D electron or hole gases with different mobilities are formed.

As the $n_i^H$ contours are distorted when $V_{TG,BG} < 0$ or $V_{TG,BG} > 0$ in comparison to $n_i$ so is the Hall mobility in comparison to the effective mobility (defined via $n_i$). Despite this fact the Hall mobility contours behave in similar fashion as the effective electron mobility contours of Ref. [4]. Electron and hole mobility decrease when the carriers are pushed away from the symmetric voltage condition $V_{BG} = V_{TG}t_{BOX}/t_{OX}$, which mainly follows from phonon and surface roughness scattering and is sometimes referred as the volume inversion (or accumulation) effect [3].

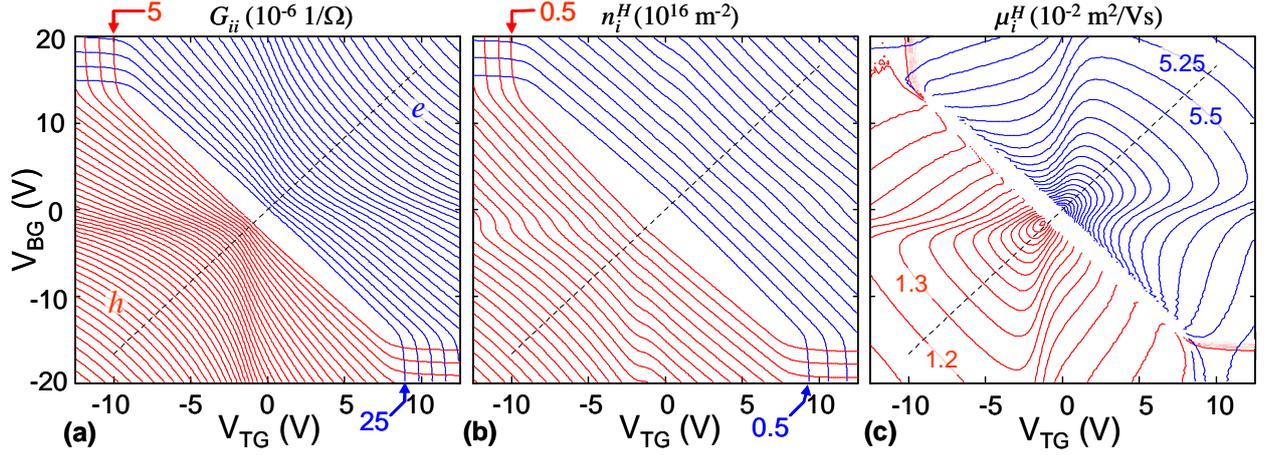

**Fig. 3:** (Color online) Experimental electron and hole (a) conductances, (b) Hall densities and (b) Hall mobilities as a function of top gate voltage $V_{TG}$ and back gate voltage $V_{BG}$. The dashed diagonal line is the symmetric gate bias line $V_{BG} = V_{TG} t_{BOX}/t_{OX}$. Blue (red) contours correspond to electrons (holes). Contours are in units given on top of each panel. Contour spacing for $G_{ee}$ and $G_{hh}$ are 25 and 5, respectively. For $n_e^H$ and $n_h^H$ the spacing is 0.5 and for $\mu_e^H$ ($\mu_h^H$) the spacing is 0.25 (0.1). The levels of few contours are indicated and other levels can identified on the basis of the contour spacing.

The electron and hole variables in Fig. 3 are simultaneously non-zero in the upper left and lower right corner. These gate voltage windows correspond to large asymmetric gate voltage ($V_{BG} \sim -V_{TG} t_{BOX}/t_{OX}$) that overcomes the Si energy gap enabling the EH bi-layer in the Si well [Fig. 1(c)]. In these bi-layer regimes $G_{ii}$ and $n_i^H$ depend weakly on the voltage of the furthest gate electrode due to screening of the other layer. Same applies for the mobility. If we inspect the $V_{TG}$ dependency of $\mu_e^H$ ($\mu_h^H$) around the threshold of hole (electron) layer close to $V_{BG} = \pm 20$ V we see no measurable effects, which could be due to entrance of the other carrier system inside the Si well and inter-layer friction.

The inter-layer effects can be brought out by inspecting the EH drag resistance $R_{he}$. Figure 4 shows $R_{he}(n_e^H, n_h^H)$ at balance Hall density $n_e^H = n_h^H$. Note that $R_{he}$ is positive and exhibits the usual monotonic decrease as a function of the layer densities.[12] It has a quite large absolute magnitude in comparison to the low temperature results obtained from compound semiconductor based bi-layers [7,8,9] and room-temperature drag effect observed between 3D gate electrode and 2D channel of Si FET [13]. However, still the intra-layer scattering is much stronger than the inter-layer friction [$R_{ii} \gg R_{ij}$ ($i \neq j$)] as expected on the basis of the mobility measurements. The effective well width for both layers is in the sub-10 nm range [see Fig. 1(c)], which leads to so large surface roughness and phonon scattering rate [3] that they simply dominate over the inter-layer momentum transfer. The intra-layer scattering is particularly strong in the hole layer. In the EH bi-layer regime the hole mobility is only ~$0.6\times10^{-2}$ m$^2$/Vs (for electrons ~$4.5\times10^{-2}$ m$^2$/Vs).

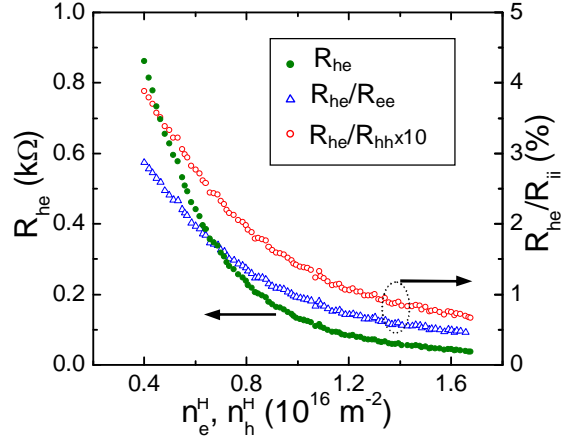

**Fig. 4:** (Color online) Left axis: EH drag resistance $R_{he}$ as a function of balanced Hall density $n_e^H = n_h^H$ in the lower right corner of Fig. 3(b). Right axis: ratios $R_{he}/R_{ee}$ and $R_{he}/R_{hh}$.

The most straightforward explanation of the inter-layer friction is given by direct Coulomb interaction between the layers. However, disorder, phonons and plasmons can strongly affect the drag.[9] Here the extremely low mobility of the hole layer makes a quantitative theoretical analysis virtually impossible, but the problem can be tackled by using phenomenological models.[14] Further complications arise from large thermal broadening, which unavoidably leads to many occupied 2D sub-bands that belong to different k-space band. Theoretical analysis of the present system is beyond the scope of this paper and it will be left for future investigations.

In summary, we have explored room temperature electron and hole transport in single 22 nm-thick double-gated Si quantum well. Our device geometry provided good independent contacts to the electron and hole systems inside the well. We investigated conductance, Hall density and Hall mobility in a broad double-gate voltage window.(Fig. 3) When the gate voltage asymmetry is not too large only either electrons or holes occupy the Si well. In these single-layer regimes the Hall mobility vs. gate voltage behavior was identified to arise from carrier distribution (position and width) dependency of surface roughness and phonon scattering. At strongly asymmetric double-gate voltage we could overcome the Si energy gap and as a result an electron-hole (EH) bi-layer was formed inside the well, which was detected as a simultaneously non-zero conductance, density and mobility in both carrier systems. The EH drag resistance was measured at balanced electron and hole densities. The drag resistance decreased monotonically from 860 Ω to 37 Ω when the electron and hole density was varied between ~0.4−1.7x10$^{16}$ m$^{-2}$.(Fig. 4)

The authors wish to thank H. Lipsanen for providing the magnet system utilized in the Hall measurements. M. Markkanen is acknowledged for technical assistance in the sample fabrication. This work has been partially funded by EU (IST-SUBTLE).